\documentclass[9pt,twocolumn,twoside]{osajnl}

\journal{jocn} 

\setboolean{shortarticle}{false}

\usepackage{textcomp} 
\usepackage{mhchem} 

\title{Cross-grating phase microscopy for nanophotonics}

\author[1,*]{Guillaume Baffou}
\affil{Institut Fresnel, CNRS, Aix-Marseille University, Centrale Marseille, Marseille, France}

\affil[*]{guillaume.baffou@fresnel.fr}

\begin{abstract}
Quantitative phase microscopies (QPMs) have been mainly used for applications in cell biology, for around 2 decades. In this article, we show how cross-grating phase microscopy (CGM), a high-resolution, high-sensitivity QPM, recently expanded the scope of QPMs to applications in nanophotonics. In particular, this article explains how the intensity and phase images acquired by CGM can be processed to determine all the optical properties of imaged nanoparticles, 2D-materials and metasurfaces. We also explain how CGM can be used as a temperature microscopy technique. This latter imaging modality led to a large variety of works in the 2010s based on the optical heating of plasmonic nanoparticles for photothermal studies in physics, chemistry and biology at the microscale, in which label-free, microscale temperature measurements were pivotal. 
\end{abstract}

\setboolean{displaycopyright}{true}

\begin{document}

\maketitle

\section{Introduction}

In the scalar approximation, a monochromatic light beam can be described as an electric field amplitude that reads in each location of space $\mathbf{r}$ and time $t$: $E(\mathbf{r,t})=\sqrt{I(\mathbf{r})}\exp(i\phi(\mathbf{r}))\exp(-i\omega t)$. While conventional imaging techniques only retrieve the intensity of a light beam $\langle|E(\mathbf{r},t)|^2\rangle=I(\mathbf{r})$, quantitative phase microscopies (QPMs), or quantitative phase imaging (QPI) techniques, enable the mapping of the phase $\phi(\mathbf{r})$ of the light beam. Since the early 2000s, several QPMs have been developed, e.g., digital holography microscopy (DHM) \cite{AO38_6994,OL30_468}, spatial light interference microscopy (SLIM) \cite{OE19_1016,AOP13_353},  diffraction phase microscopy (DPM) \cite{AOP6_57,AO46_A52}, and cross grating phase microscopy (CGM) \cite{OA39_5715,OE17_13080,JPDAP54_294002}  among others. These last two decades, the main domain of application of QPMs has been live cell imaging \cite{S13_4170,OLE135_106188}. 

In this article, we show how cross-grating phase microscopy (CGM), a high-resolution, high-sensitivity QPM, recently expanded the scope of QPMs to applications in nanophotonics. In particular, this article explains how the intensity and phase images acquired by CGM can be processed to determine all the optical properties of imaged nanoparticles, 2D-materials and metasurfaces. We also explain how CGM can be used as a temperature microscopy technique. This latter imaging modality led to a large variety of works in the 2010s based on the optical heating of plasmonic nanoparticles for photothermal studies in physics, chemistry and biology at the microscale, in which label-free, microscale temperature measurements were pivotal. 

\section{Cross-grating phase microscopy (CGM)}

\subsection{Principle of CGM}

\begin{figure*} [ht]
\begin{center}
\includegraphics{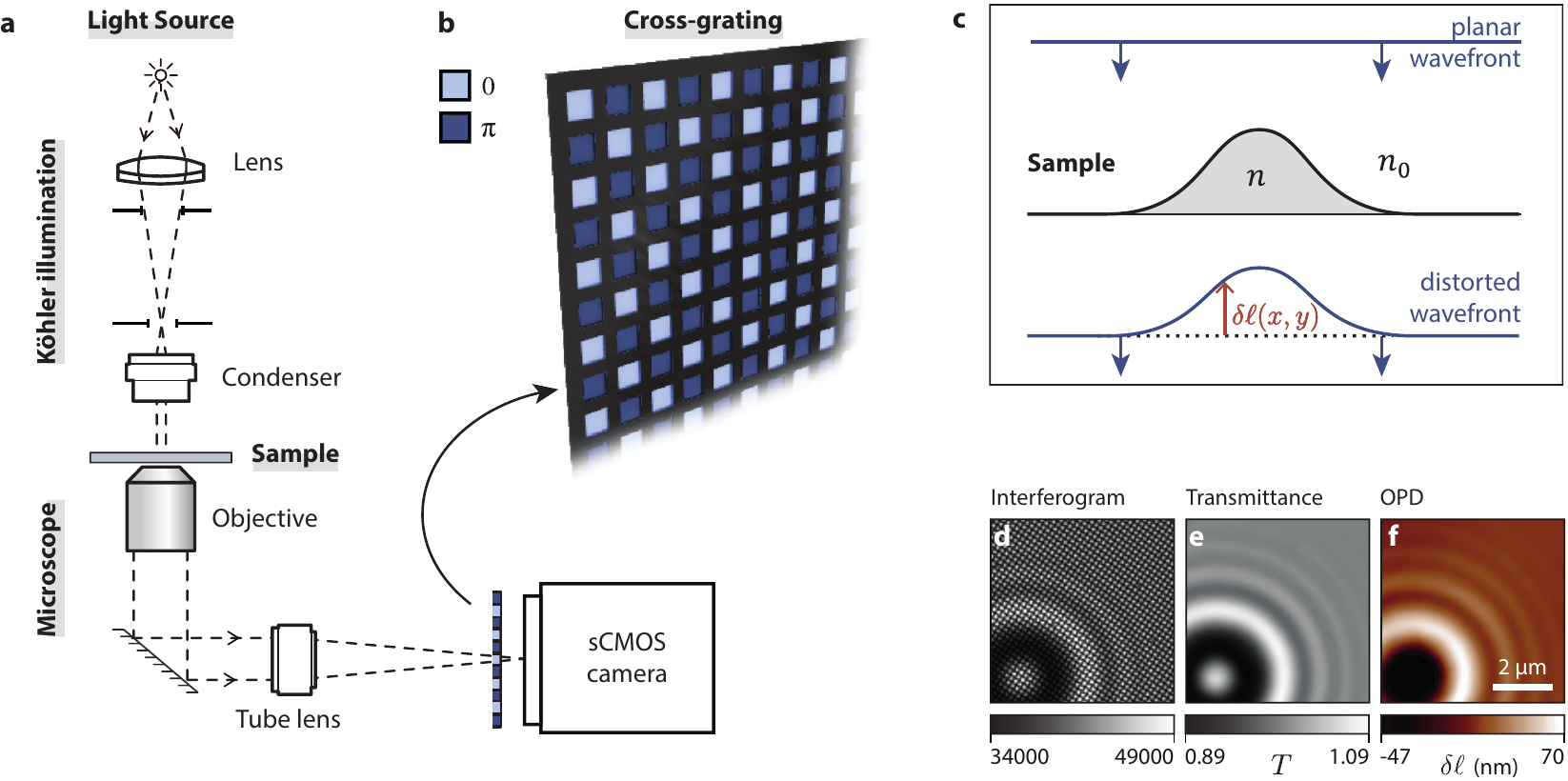}
\end{center}
\caption{ \label{setup} 
(a) Typical cross-grating microscopy (CGM) setup. (b) QLSI cross-grating. (c) Schematic of the wavefront distortion created by an imaged object that is measured by CGM. (d) Raw image acquired by the camera (interferogram), featuring a 2-\textmu m dielectric bead. (e) Transmittance and (f) optical path difference (OPD) processed from the interferogram. Panels reproduced with permission from : (a,b) Ref. \cite{ACSP8_603}, American Chemical Society; (d,e,f) Reproduced with permission from Ref. \cite{JPDAP54_294002}, IOP Publishing.}
\end{figure*}

Cross-grating wavefront sensing is an imaging technique that can map the wavefront profile of a light beam \cite{OA39_5715,OE17_13080,JPDAP54_294002}. It is based on the association of a specific 2-dimensional (2D) diffraction grating, aka cross-grating \cite{book_BornWolf}, and a regular camera, separated by a millimetric distance (Fig. \ref{setup}a). The postprocessing of the measured image (called an interferogram) enables the retrieval of both the intensity $I$ and the wavefront profile $W$ of a light beam. One particular instance of cross-grating wavefront sensing, introduced in 2000, generally called quadriwave lateral shearing interferometry (QLSI or QWLSI), uses a $0-\pi$ checkerboard cross-grating  (Fig. \ref{setup}b) \cite{OA39_5715}. In 2009, just because the wavefront profile can be simply related to the phase of the light beam by the relation $\phi=2\pi W/\lambda$, it was demonstrated that QLSI could also be implemented on a microscope to be used as a QPM technique, with applications in cell biology \cite{OE17_13080}. In the following, although all the applications in biology and nanophotonics have been done using QLSI so far, we will prefer the more general appellation cross-grating wavefront sensing, or rather cross-grating phase microscopy (CGM) over QLSI, for the sake of simplicity and generality.

When implemented on a microscope, CGM measures an intensity image that equals the transmittance $T$ of the sample, and a wavefront profile $W$ that equals the optical path difference (OPD) $\delta\ell$ experienced by the probe beam crossing the sample: 
\begin{eqnarray}
\delta\ell(x,y)&=&\int(n(x,y,z)-n_0)\mathrm{d}z\\
T(x,y)&=&I(x,y)/I_0(x,y)
\label{eq:deltaell}
\end{eqnarray}
where $n$ is the refractive index of the imaged object, $n_0$ the refractive index of the surrounding medium, and $I_0$ the intensity image created by the illumination only (without the refractive object).

\subsection{Advantages and limitation of CGM}

CGM is a common-path QPM technique, meaning that no reference beam has to be implemented on the microscope. This particularity is a great advantage because CGM can be simply implemented on any microscopy, just like a regular camera, and because reference beams are likely to affect the quality of the image (fringes, instabilities, ...).

Also, CGM is achromatic. It can be used with incoherent, broad band light sources. This feature avoids the presence of unwanted fringes, and offers the possibility to easily vary the illumination wavelength using a monochromator.

The spatial resolution of CGM is around $10\times$ better than the most famous wavefront sensing technique, namely Shack-Hartmann wavefront sensing. CGM still features a reduction by a factor of 3 of the spatial resolution dictated by the pixel density of the camera. This is the price to pay to have the information on both the intensity $I$ and the phase $\phi$ of light beam in a single interferogram image. However, if need be, it is still possible to achieve a diffraction limited spatial resolution with CGM, by oversampling the image with a high magnification objective \cite{OE17_13080}. For instance, at an illumination wavelength of $\lambda=550$ nm, with a $100\times$, 1.3 NA (numerical aperture), objective and camera pixels (dexels) of 6.5 \textmu m, the pixels of the image represent of size of 65 nm at the object plane. The reduction by a factor of 3 leads to an effective phase pixel size $d=195$ nm. The Niquist criteria states that spatial frequencies $f=1/P$ that can be captured range from $f/2$, corresponding in this case to $P=390$ nm. This value is better than the spatial resolution imposed by the microscope $1.22\lambda/\mathsf{NA}=516$ nm (this latter expression considers a zero-NA illumination). As a consequence, this reduction by a factor of 3 is usually not a restriction when using $100\times$ objectives.

\section{Optical characterisation of\newline nanoparticles by CGM}

\begin{figure} [ht]
\begin{center}
\includegraphics{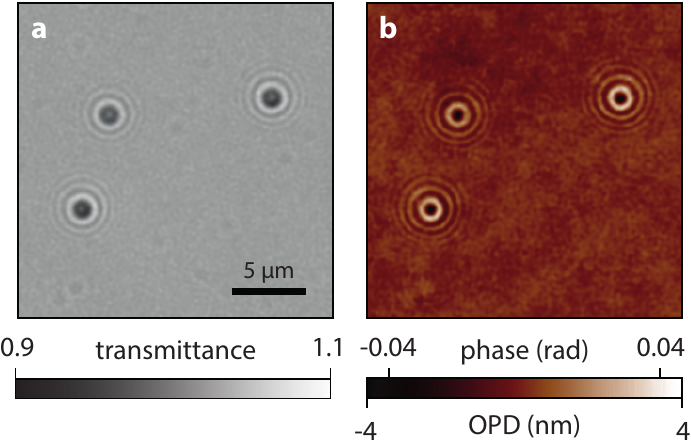}
\end{center}
\caption{ \label{nanoparticles} 
(a) Transmission image of 100-nm gold nanoparticles acquired by CGM. (b) Associated OPD/phase image. Experimental conditions: $\lambda=540$ nm, $100\times$ objective.}
\end{figure} 

Figure \ref{nanoparticles} presents transmittance and OPD measurements of isolated 100-nm gold nanoparticles deposited on a glass substrate in air, acquired by CGM. Due to light propagation through the microscope and diffraction, the image of such small objects looks like an Airy pattern, with ring sizes directly related to light wavelength, collection NA and focus. With sub-diffraction objects like here, equation \eqref{eq:deltaell} does not hold anymore. It may be the reason why imaging nanoparticles have not been investigated by QPM until very recently: the nature of the information that can be retrieved from the image is not obvious. Interestingly, it was demonstrated that these transmission $T$ and OPD $\delta\ell$ images of a nanoparticle contain precious information. In particular, the complex optical polarizability $\alpha$ of an imaged nanoparticle can be retrieved \cite{O7_243}. Defined by $\mathbf{p}=\varepsilon_0\alpha\mathbf{E}_0$ where $\mathbf{p}$ is the polarisation vector of the nanoparticle and $\mathbf{E}_0$ the local electric field amplitude at the nanoparticle location, the complex optical polarizability $\alpha$ can be determined by combining the transmission and OPD images this way:
\begin{equation} 
\alpha=\frac{i\lambda n_0}{\pi}\iint\left(1-\sqrt{T(x,y)}e^{i2\pi\delta\ell(x,y)/\lambda}\right)\mathrm{d}x\mathrm{d}y
\label{eq:alpha}
\end{equation}
The integral amounts to a pixel summation over the Airy pattern. This quantity is of utmost importance because, for dipolar nanoparticles, it can be used to compute all the optical cross sections of the nanoparticle (extinction, absorption and scattering):
\begin{eqnarray}
\sigma_{ext}&=&\frac{k}{n_0}\mathrm{Im}(\alpha)\\
\sigma_{abs}&=&\frac{k^4}{6\pi}|\alpha|^2\\
\sigma_{sca}&=&\sigma_{ext}-\sigma_{abs}
\end{eqnarray}
where $k=2\pi/\lambda$ is the free-space wave vector of light. Usually, the determination of these three quantities requires three different experimental setups. Here, with one single CGM interferogram image, they can be determined all at once. Khadir et al. reported this CGM application in 2020, illustrated on gold and polystyrene nanoparticles \cite{O7_243}. A major strength of the technique is that, although the $T$ and $\delta\ell$ images are highly dependent on both the focus and the numerical aperture of the microscope, Equation \eqref{eq:alpha} isn't, making the measurements particularly precise and reliable.

Bon et al. recently developed a practical application of imaging metal nanoparticles using CGM. Just because the OPD image of a nanoparticle features a sharp contrast variation at the focus when moving the sample in the optical axis direction, imaging of metal nanoparticles was proposed as a means to finely adjust and maintain the focus of a microscope, leading to a new kind of autofocus system with applications in biomicroscopy \cite{NM15_449,US10261305B2}.

\section{Optical characterisation of 2D-materials by CGM}
Just like the optical response of nanoparticles (0D object) is characterised by the complex optical polarizability, the optical response of 2D materials is characterised by its complex optical conductivity $\sigma_\mathrm{2D}$, defined by $\mathbf{J}_\mathrm{2D}=\sigma_\mathrm{2D}\mathbf{E}$, where $\mathbf{E}$ is the total electric field within the material, and $\mathbf{J}_\mathrm{2D}$ the 2D electronic current density. A very similar expression, mixing again the transmission and OPD images, can be used to determine $\sigma_\mathrm{2D}$ \cite{ACSP4_3130}:
\begin{equation}
\sigma_\mathrm{2D}(x,y)=\varepsilon_0c(n_1+n_2)\left[\frac{1}{\sqrt{T(x,y)}e^{i2\pi\delta\ell(x,y)/\lambda}}-1\right]\label{eq:sigma2D}
\end{equation}
where $n_1$ and $n_2$ are the refractive indices of the upper and lower media and $c$ the speed of light in vacuum. Note that this Equation \eqref{eq:sigma2D} does not involve an integral, unlike Equation\eqref{eq:alpha}, meaning that $\sigma_\mathrm{2D}$ is also an image. CGM enables the mapping of the optical conductivity $\sigma_\mathrm{2D}(x,y)$ throughout the 2D-material.

\begin{figure} [ht]
\begin{center}
\includegraphics{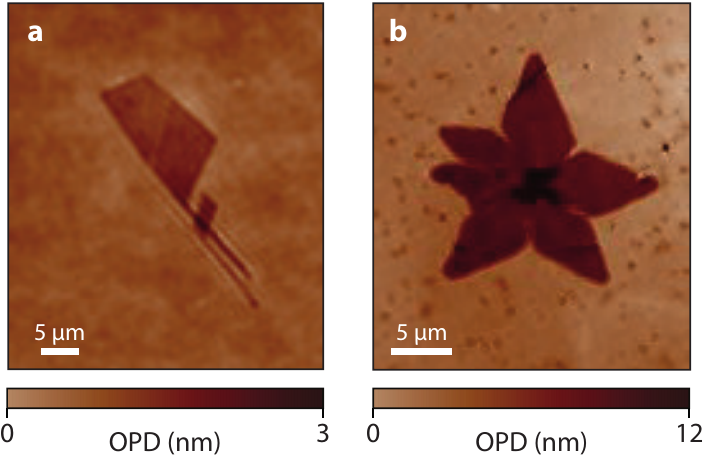}
\end{center}
\caption{ \label{2Dmat} 
OPD images of (a) a graphene structure composed of a single layer near a double layer and (b) a \ce{MoS2} structure, acquired by CGM. Reproduced with permission from Ref. \cite{ACSP4_3130}, American Chemical Society.}
\end{figure}

This optical metrology technique using CGM was introduced by Khadir et al. in 2017 \cite{ACSP4_3130}. The principle was illustrated on graphene and molybdenum disulfide (\ce{MoS2}) (Figure \ref{2Dmat} shows OPD images of graphene and \ce{MoS2} measured by CGM, reproduced from Ref. \cite{ACSP4_3130}).

Just like with nanoparticles, the expression \eqref{eq:deltaell} of $\delta\ell$ just based on refractive indices considerations may be inappropriate because, here again, the imaged objects are spatially confined at the nanometric scale. In particular, for graphene, which is only one carbon atom thick, the refractive index has no meaning anymore because a refractive index is only defined for a bulk material. It would make more sense to always deal with the 2D optical conductivity $\sigma_\mathrm{2D}$, which does not suffer from this limitation. However, the community is still attached to the use of refractive indices for 2D materials. For graphene, a refractive index can still be defined provided an arbitrary thickness is chosen, which is normally considered to be the interlayer distance in graphite, around 0.33 nm. Only when considering this arbitrary graphene thickness, one can determine an effective refractive index (around 2.5). For multi-atomic-layer 2D-materials, such as \ce{MoS2}, dealing with refractive indices is less a problem.

\section{Optical characterisation of metasurfaces by CGM}
\begin{figure*} [ht]
\begin{center}
\includegraphics{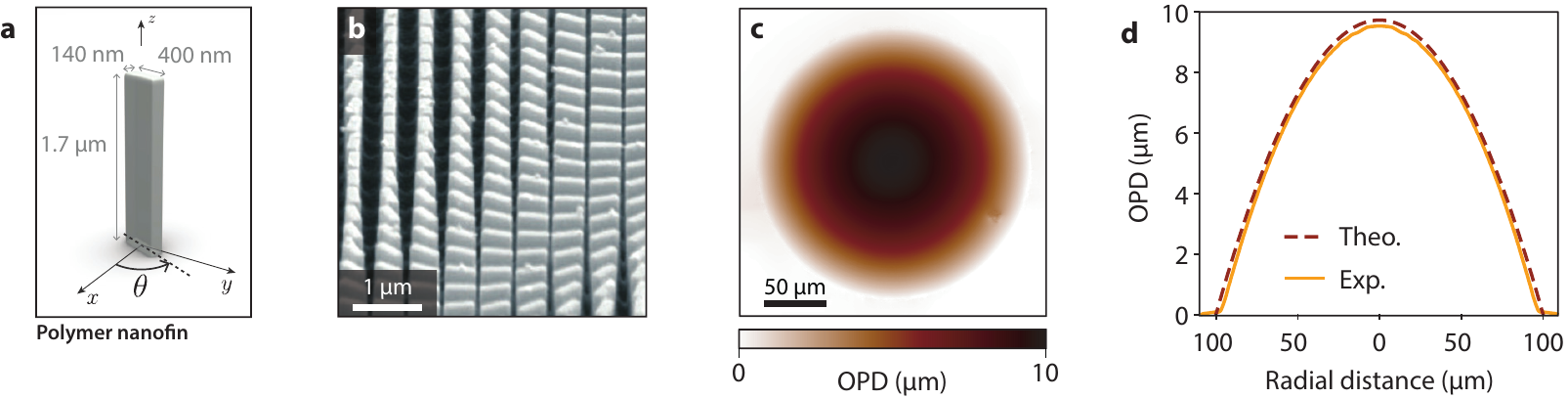}
\end{center}
\caption{ \label{metasurfaces} 
(a) Geometry of the metaunit, a polymer nanofin, composing the metalens. (b) SEM image of a part of the metalens. (c) OPD image of a metalens recorded by QLSI along with its (d) radially averaged profile. Reproduced with permission from Ref. \cite{ACSP8_603}, American Chemical Society.}
\end{figure*}

Besides nanoparticles and 2D-materials, another category of trendy objects in nanophotonics are metasurfaces. A metasurface is a non-uniform array of nanostructures (called meta-units) that are densely distributed so that their optical properties can be considered as non-uniform, but continuous \cite{NRM5_604,RPP81_026401,S358_eaam8100}. Meta-units can be engineered to locally modify the intensity and phase of a light beam passing through. Thus, using CGM as a characterisation tool of metasurfaces seems natural. This application of CGM has been demonstrated in 2021 on two types of metasurfaces, namely Pancharatman-Berry and effective-refractive-index types \cite{ACSP8_603}. In particular, CGM was used to characterize metalenses (metasurfaces producing an hyperbolic phase profile to focus a light beam). Instead of just looking at a focused spot at the image plane of the metalens, as commonly done, CGM enables the direct mapping of the phase profile at the metalens plane, highlighting and quantifying all the possible aberrations and origins of an imperfect point-spread-function. Figure \ref{metasurfaces} displays measurements of OPD images acquired on a Pancharatnam-Berry metalens, composed of nanofins metaunits.

\section{Microscale temperature imaging by CGM}
\begin{figure*} [ht]
\begin{center}
\includegraphics{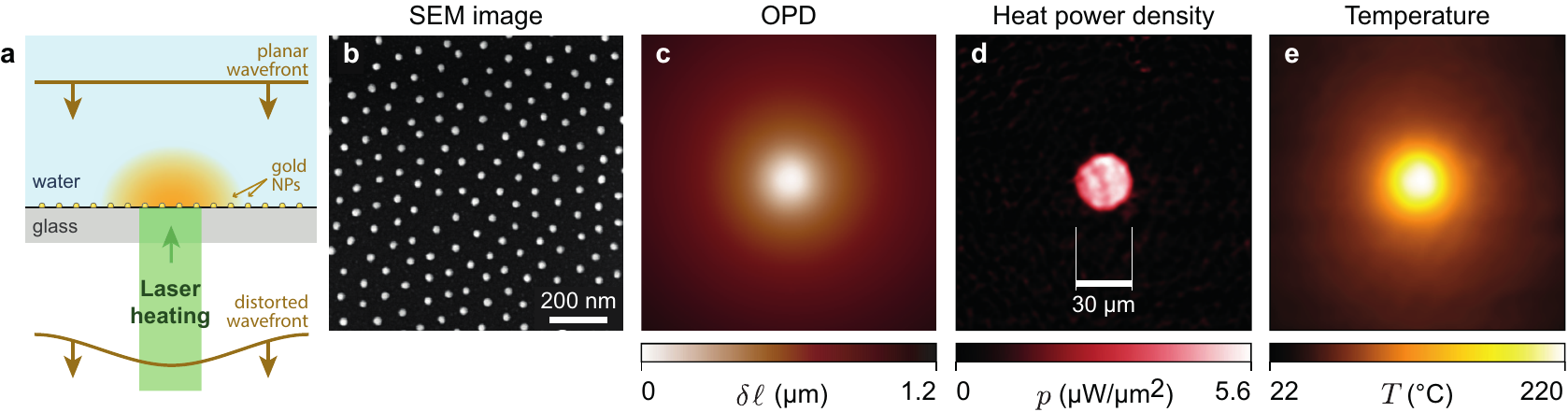}
\end{center}
\caption{ \label{temperature} 
(a) Schematic of the system: a gold nanoparticle (NP) layer (not to scale) heated with a laser at $\lambda=532$ nm, and an incoming light wavefront distorted by the temperature gradient in the liquid. (b) Scanning electron microscopy (SEM) image of the gold nanoparticle layer. (c) OPD image resulting from the temperature gradient. (d) Postprocessed heat source density from the OPD (c). Post-processed temperature distribution from the heat power density (d). Reproduced with permission from Ref. \cite{JPCC118_4890}, American Chemical Society.}
\end{figure*}

In 2012, the ability of CGM to become a temperature microscopy technique was demonstrated \cite{ACSNano6_2452,APL102_244103}. CGM could image microscale temperature gradients in the context of application in plasmonics where gold nanoparticles were heated by a laser within the field of view of a microscope, in a liquid environment. Because the refractive index of a liquid is temperature dependent, any microscale temperature gradient in a liquid results in a thermal lens effect, or equivalently a mirage effect, distorting an incoming probe light beam (Fig. \ref{temperature}a). This wavefront distortion can be mapped by CGM and postprocessed using an inversion algorithm to retrieve the 3D temperature distribution in the liquid from the OPD image, measured by CGM (Fig. \ref{temperature}e). Interestingly, this approach also enables the mapping of the microscale heat source density (power per unit area) delivered by the heat source (Fig. \ref{temperature}d).

Following this imaging modality of CGM, a large variety of photothermal effects at the microscale could be studied for a decade, most of them arising from the laser heating of gold nanoparticles (following the sample geometry of Figure \ref{temperature}a,b), namely absorption spectroscopy of nanoparticles \cite{PRB86_165417}, laser-controlled adhesion of living cells \cite{ACSNano6_7227}, collective photothermal effects with gold nanoparticle arrays \cite{ACSNano7_6478}, super-heating of fluids \cite{JPCC118_4890}, micro-bubble generation around gold nanoparticles \cite{JPCC118_4890}, microscale temperature shaping at will \cite{N6_8984,SR9_4644}, photothermal properties of metallic nanowire networks \cite{ACSNano9_5551},  solvothermal chemistry \cite{ACSOmega1_2}, heat-shock protein expression in living cells \cite{S14_1801910}, photothermal effects in gold nanoholes \cite{ACSP6_1763,NL12_8811,APLP6_101101} and microscale thermophoresis in liquid \cite{NL12_8811,JPCC125_21533}.

Other techniques aiming at mapping microscale temperature gradients exist \cite{book_Baffou,N4_4301,NT19_126,JPPC30_2}. They are often based on fluorescence measurements, for instance within living cells or around plasmonic nanoparticles. As soon as fluorescent molecules are dispersed in the medium of interest, a fluorescence map gives a temperature map provided a proper fluorescence-temperature calibration was established. The issue with fluorescence-based approaches is two-fold: (i) It is invasive. One needs to modify the sample (add fluorescent compounds), which is not always possible. (ii) It is not reliable as fluorescence is also depending on many other factors (especially in living cells) and can be affected by thermo-bleaching and photobleaching \cite{NM11_899,NM12_801,NM12_802,NM12_803}. Using CGM instead of fluorescence microscopy, i.e., monitoring refractive index variations instead of fluorescence variations, solves all the above-mentioned problems.

\section{Conclusion}
This article reviews the recent examples of QPM studies for nanophotonics, all of them conducted by cross-grating phase microscopy (CGM), a high-resolution, high-sensitivity quantitative phase microscopy. The high-sensitivity, that can reach a fraction of a hydrogen atom in wavefront reconstruction \cite{JPDAP54_294002}, certainly explains why CGM could successfully characterize small objects, such as nanoparticles or one-atom thick materials.

To date, quantitative phase microscopy techniques have been mainly used by biologists, typically to image living cells in culture. Yet, the phase of light is first of all a physical quantity, and one may be surprised physicists do not exploit that much the ability to map the phase of a light beam. We predict that this paradigm is about to change. Nanophotonics, in particular, is full of practical cases where mapping the phase of light would provide a wealth of information.


\end{document}